\documentclass[a4paper,11pt]{article}

\usepackage{a4wide}
\usepackage[english]{babel}
\usepackage{amsmath,amssymb,amsfonts,amsthm}

\newcommand{\vol}{\int d^4x \sqrt{g}}
\newcommand{\volbar}{\int d^4x \sqrt{\bar g}}
\newcommand{\sqrtbar}{\sqrt{\partial^\mu\bar \phi \partial_\mu\bar\phi +\lambda^2}}

\title{Preferred foliation effects in Quantum General Relativity}
\author{
Tim Koslowski\thanks{e-mail: \texttt{tkoslowski@perimeterinstitute.ca}}\\
\hfill\\
 Perimeter Institute for Theoretical Physics\\
  31 Caroline St. N, ON N2L 2Y5, Waterloo, Canada \\
\hfill\\
  Alexander Schenkel\thanks{e-mail: \texttt{aschenkel@physik.uni-wuerzburg.de}}\\
  \hfill\\
  Institut f\"ur Theoretische Physik und Astrophysik\\
    Universit\"at W\"urzburg\\
  Am Hubland, 97074 W\"urzburg, Germany}

\newcommand{\hor}{Ho\v{r}ava}

\begin{document}

\maketitle

\begin{abstract}
We investigate the infrared (IR) effects of Lorentz violating terms in the gravitational sector using
 functional renormalization group methods similar to Reuter and collaborators \cite{Reuter:1996cp}.
 The model we consider consists of pure quantum gravity coupled to a preferred foliation, described
 effectively via a scalar field with non-standard dynamics. We find that vanishing Lorentz violation is a UV
 attractive fixed-point of this model in the local potential approximation. Since larger truncations may lead to
 differing results, we study as a first example effects of additional matter fields
 on the RG running of the Lorentz violating term and provide a general argument why they are small.
\end{abstract}

\section{Introduction and motivation}

One of the major unsolved problems of theoretical physics is the construction of a physically acceptable
and predictive UV completion of quantum gravity. The perturbative nonrenormalizability of gravity hints
that its quantization requires new physics, such as a preferred foliation modeled, e.g., in Einstein-aether
theory \cite{Eling:2004dk}. Preferred foliations gained a lot of attention recently after \hor's proposal
\cite{Horava:2009uw} of a perturbatively renormalizable quantum theory of gravity. \hor's model is a higher
derivative gravity theory that avoids  troublesome unitary ghosts by retaining second derivatives in time while
allowing higher derivatives in space. The preferred foliation can be encoded \cite{Germani:2009yt} in a
St\"uckelberg-type field $\phi$ which defines an irrotational unit time-like vector field
$n_\mu=\phi_{,\mu}/\sqrt{g^{\nu\rho}\phi_{,\nu}\phi_{,\rho}}$, where the preferred foliation is defined by
constant $\phi$ surfaces. The extrinsic curvature of the preferred foliation is
$K_{\mu\nu}=\mathcal L_n(g_{\mu\nu}+n_\mu n_\nu)$, where $\mathcal L_n$ denotes the Lie derivative
in direction of $n$. The low energy limit of \hor's theory contains in addition to the usual Einstein-Hilbert
action also the Lorentz violating action given by
\begin{flalign}
\label{eqn:LVaction}
S_\mathrm{LV} =\frac{b}{16\pi G_N}\int d^4x\sqrt{\vert g\vert} ~ K^2 +~\text{higher derivative terms}~,
\end{flalign}
where $G_N$ denotes Newton's constant and $K=g^{\mu\nu}K_{\mu\nu}$. $b$ is a new coupling constant
 that vanishes in general relativity. A possible $K^{\mu\nu}K_{\mu\nu}$-term is absorbed in \hor's definition
 of the speed of light. The modification (\ref{eqn:LVaction}) is a low energy effective field theory,
describing the effects of quantum gravity at a scale $k$ where higher derivative terms are negligible.
In order to bring this theory into contact with observation we use the renormalization group
equation to evolve this action to IR scales at which experimental tests for Lorentz violation are being performed.

A Wilsonian approach to the renormalization group equation for (\ref{eqn:LVaction}) leads to considering
 nonperturbative renormalization of Einstein-aether theories, which is very involved. A much simpler
 on-shell equivalent to low energy \hor~theory has been found in \cite{Afshordi:2009tt} in terms of the cuscuton field:
\begin{flalign}
\label{eqn:cuscaction}
S_\mathrm{cusc}=\int d^4x \sqrt{\vert g\vert} \Bigl( \sqrt{g^{\mu\nu} \phi_{,\mu} \phi_{,\nu}  } - \frac{a_1}{2}  \phi^2  \Bigr) ~.
\end{flalign}
This model generates the $K^2$ term dynamically and, given sufficient boundary conditions, leads to the
 dynamical emergence of a constant mean curvature foliation. The quadratic cuscuton coupling $a_1$
  (which has dimension $\textrm{mass}^{-2}$ in our normalization) is related to the Lorentz violating parameter
  $b$ by
\begin{flalign}
\label{eqn:mum}
a_1=\frac{8\pi G_N}{b}~.
\end{flalign}
The observational constraints on the gravitational Lorentz violating parameter $b$ are much weaker than the
 bounds on Lorentz violating parameters in extensions of the standard model. A cosmological bound constraining
 $-b=0.003\pm0.014 $ at $95\% ~\text{CL}$ can be found in \cite{Afshordi:2006ad}. Notice that the Lorentz
  violating parameter $b$ is dimensionless, so there is no power counting argument for its running.

The advantages of Reuter's programme, which achieves to be at the same time nonperturbative and background
independent, suggest to use this Wilsonian framework for the investigation of Lorentz violating models. As a start
 into the Wilsonian investigation of Lorentz violating models of this type, we will consider an adapted version of the
  cuscuton model (\ref{eqn:cuscaction}) generalized to an arbitrary even power-series potential $V(\phi)$ and discuss
  the $\beta$-function of the quadratic cuscuton coupling in presence of pure quantum gravity. The main result is that in the local potential approximation
  the $\beta$-function of the Lorentz violating parameter $b$ has a UV attractive Gaussian fixed-point $b=0$, i.e.~the
   $\beta$-function of the absolute value $\vert b\vert$ is negative in the vicinity of this fixed-point. Fortunately, the
   corresponding IR growth of $b$ is bounded due to an IR decreasing $\beta$-function $\beta(b)$, if we assume a
    sufficiently semi-classical RG flow in the gravitational sector.  We show that coupling additional matter fields to this
     theory has only minor effects on the RG flow of the Lorentz violating parameter, at least for small matter couplings
     and Lorentz violations $b$. Notice that since the cuscuton model differs from \hor's theory, both through off-shell
      terms and higher derivative contributions, one can not apply these results to \hor's theory itself, but one should
      consider them as a toy model to gain a first insight.

The outline of this paper is as follows: In section \ref{sec:FRGE} we provide the background on Wetterich's
 functional renormalization group equation \cite{Wetterich:1992yh} necessary in order to be self-contained.
 Section \ref{sec:general}  explains the basic idea of IR attractive gauge symmetries using a toy model.
 In section \ref{sec:cuscuton} we calculate explicitly the flow of the parameter $b$ within the cuscuton model
 coupled to quantized gravity. The introduction of matter to this model is discussed in section \ref{sec:matter}.
 We give an outlook towards further investigations in this direction in section \ref{sec:conc}.

\section{\label{sec:FRGE}The functional renormalization group equation (FRGE)}

In this section we present a brief review of Wetterich's functional renormalization group equation (FRGE)
\cite{Wetterich:1992yh} to provide the prerequisites for this paper using a Euclidean scalar field theory;
detailed introductions can e.g.~be found in \cite{Pawlowski:2005xe}\cite{Gies:2006wv}\cite{Bagnuls:2000ae}\cite{Morris:1998da}
and reviews on the application to gravity in \cite{Niedermaier:2006wt}\cite{Codello:2008vh}\cite{Percacci:2007sz}.

The starting point for formulating the FRGE is the partition function defined by the regularized Euclidean path-integral
\begin{flalign}\label{equ:W-definition}
 \exp(W_k[J])=\int \mathcal{D}_\Lambda\chi~e^{-S[\chi]-\Delta S_k[\chi] + J.\chi}~,
\end{flalign}
where $\Lambda$ denotes an overall UV cutoff, $S$ the bare action (at the cut-off) and $\Delta S_k$ an
 IR regulator that suppresses momentum modes with $p^2 < k^2$ by giving them a mass of order $k$
 while effectively vanishing for modes with $p^2 > k^2$. A typical suppression term is given by
\begin{flalign}
 \Delta S_k[\chi] = \frac{1}{2} \int d^4x~ \chi(x) R_k(-\square) \chi(x) = \frac{1}{2}\int \frac{d^4p}{(2\pi)^4} ~\tilde\chi(p) R_k(p^2) \tilde\chi(-p)~.
\end{flalign}

The effective average action $\Gamma_k$ is defined via the Legendre transformation and subsequent subtraction
of the IR regulator
\begin{flalign}\label{equ:Gamma-definition}
  \Gamma_k[\phi] := \int d^4x J(x) \phi(x) - W_k[J]- \Delta S_k[\phi]~,
\end{flalign}
where $\phi(x) = \langle \chi(x) \rangle = \delta W_k[J] / \delta J(x)$ is the expectation value of the field $\chi$.
To calculate the scale derivative $\partial_t:=k\partial_k$ of $\Gamma_k$ we take the scale derivative of
(\ref{equ:Gamma-definition}), substitute $W_k$ with (\ref{equ:W-definition}) and reexpress the RHS in terms
of variations of $\Gamma_k$, resulting in Wetterich's FRGE
\begin{flalign}
\label{eqn:wetterich}
 \dot\Gamma_k[\phi] = \frac{1}{2} \mathrm{Tr} \left[ \dot R_k \cdot\bigl(\Gamma^{(2)}_k[\phi] + R_k\bigr)^{-1}   \right]~,
\end{flalign}
where dot denotes $\partial_t$ and $\Gamma^{(2)}_k$ denotes the second variation of $\Gamma_k$ w.r.t.~the fields.
The trace extends over all fields.

The usual effective action $\Gamma$ is attained in the limit $k\to0$, since $\Delta S_k$ vanishes in this limit.
The usual bare action is attained in the limit $k,\Lambda\to\infty$, because $\Delta S_k$ diverges in this limit,
making the saddle point approximation for (\ref{equ:W-definition}) exact. The effective average action $\Gamma_k$
is thus an interpolation between the usual bare action $S$ (for $\Lambda\to\infty$) and the usual effective action
$\Gamma$ and effectively describes physics at the scale $k$, because its tree-level accurately describes the effects
of momentum modes above $k$. This allows one to derive an effective action without explicitly solving a path-integral
by specifying an initial condition $\Gamma_{k=\Lambda}$ and using (\ref{eqn:wetterich}) to evolve $\Gamma_k$ to
the physically relevant scale $k$.

To calculate $\beta$-functions from (\ref{eqn:wetterich}), one expands $\Gamma_k$ in terms of field monomials
in the flow equation and equates the coefficients of the field monomials on the LHS with the corresponding coefficient
on the RHS. However, the trace on the RHS of the flow equation produces in general an infinite number of field monomials,
meaning that one can not stick with an action of a particular form, but one is forced to make the most general ansatz
compatible with field content and symmetries. Using the most general ansatz for the effective average action however
yields an insurmountable calculation, so for practical purposes one will have to make an ansatz by selecting the
``most important'' monomials and expand the trace on the RHS only within this ansatz. The strength of the flow equation
is that one can use expansions that are not necessarily perturbative, in particular an expansion in the number of derivatives
 of the field operator.
 The lowest order of the derivative expansion is the local potential ansatz, which reads for a standard
 scalar field
\begin{flalign}
\Gamma_k[\phi] = \int d^4x \left( \frac{1}{2}\partial_\mu\phi \partial^\mu \phi + V_k(\phi)   \right)~.
\end{flalign}
Practically, one will consider only the lowest dimensional monomials, so imposing a $\mathbb Z_2$ symmetry,
we may consider
\begin{flalign}
V_k(\phi) = \frac{m^2_k}{2} \phi^2 + \frac{\lambda_k}{4!} \phi^4 +\dots~.
\end{flalign}
The focus of this paper is the renormalization of the cuscuton field coupled to gravity, which we will do in the local
potential approximation. For this we have to give a generalization of the FRGE methods so gravity can be included.
An important ingredient are background-field methods, as used in various applications to gauge and gravity theories.
This is particularly important in gravity, since coarse graining in a background independent theory needs a split of the
metric $g_{\mu\nu}=\bar g_{\mu\nu}+h_{\mu\nu}$ into an arbitrary background metric $\bar g_{\mu\nu}$ and
fluctuations $h_{\mu\nu}$, which are not assumed to be small.
Following Reuter \cite{Reuter:1996cp}, we use the background Laplacian to define the momentum scale $k$ of the
fluctuation field. In contrast to the perturbative graviton expansion however, one does neither assume a fixed background
nor that the fluctuations are small; it is hence not necessary to make an expansion in terms of gravitons for the price
that the effective action of gravity depends on both, $\bar g_{\mu\nu}$ and $h_{\mu\nu}$.

The trace on the RHS of the flow equation can then be expanded background independently in terms of curvature invariants
using the heat kernel expansion. The first two terms in this expansion yield the invariants occurring in the Einstein-Hilbert
action. The Einstein-Hilbert truncation with harmonic background field gauge fixing reads
\begin{multline}
\Gamma_\mathrm{EH}[\bar g,h] = 2\kappa^2 \int d^4x \sqrt{g} \bigl(-R[g]+ 2 \Lambda \bigr) \\
+ \kappa^2 \int d^4x\sqrt{\bar g} \bar D^\rho \left( h_{\mu\rho} -\frac{\bar g_{\mu\rho}}{2} h  \right) \bar g^{\mu\nu}\bar D^\sigma \left( h_{\nu\sigma} -\frac{\bar g_{\nu\sigma}}{2} h  \right) +\text{ghosts}~,
\end{multline}
where $h=\bar g^{\mu\nu} h_{\mu\nu}$ and $\kappa^{-2} = 32 \pi G_N$. We did not give the explicit form of the ghost
action, since it does not explicitly contribute to the running of the quadratic cuscuton coupling in our truncation.

The $R_k$ and $\Gamma^{(2)}_k$ terms appearing in the FRGE (\ref{eqn:wetterich}) include only the functional derivatives with
respect to the fluctuation fields. The background fields set the scale $k$ present in the regulator function, i.e.~we define for
 $h_{\mu\nu}$, as well as for all other dynamical fields, the regulator action with respect to $\bar g_{\mu\nu}$ and the
 corresponding Laplacian $\bar \square$
\begin{flalign}
\Delta S_k[\bar g, h] =\kappa^2 \int d^4x \sqrt{\bar g} ~h_{\mu\nu}~R_k^{\mu\nu\rho\sigma}(\bar g,- \bar \square) ~h_{\rho\sigma}~.
\end{flalign}
The background gauge condition is satisfied for vanishing fluctuations, so we may, after the FRGE has been derived,
set the averaged fluctuations to zero, i.e.~$h_{\mu\nu}=0$. Then the background field $\bar g_{\mu\nu} $ can be
interpreted as the averaged quantum metric field.

A useful trick \cite{Fraser:1984zb} in the evaluation of the trace on the RHS of the flow equation is to obtain a formal
expression $\frac 1 2\textrm{Tr}(\hat O[\phi])=F[\phi]$, so the flow equation reads
$\sum_i \dot{\lambda}_i O_i[\phi] =F[\phi]$, where $\lambda_i$ denotes a coupling and $O_i[\phi]$ a field monomial.
It is now useful to insert families of field configurations that (1.) project onto the truncation on the LHS, (2.) identify the
 terms in the truncation uniquely and (3.) simplify the evaluation of $F$, e.g.~constant matter fields to project onto
 a local potential ansatz or flat metrics to project onto monomials that do not involve curvature invariants.

\section{\label{sec:general}IR symmetries: General scenario}

Let us assume that we integrate out short distance physics above a very high scale $\Lambda$ and hence produce
an effective average action $\Gamma_\Lambda[\phi_i]$ with field content $\{\phi_i\}_{i\in \mathcal I}$. Let us
furthermore consider a gauge group $\mathcal G$ acting on the field content $\{\phi_i\}$, so the gauge invariant
field content is $\{\phi_i\}/\mathcal G$. The question that we want to consider is whether this particular gauge
symmetry is IR attractive or whether one needs fine-tuning to retain gauge symmetry in the IR. To answer this question
we use the exact renormalization group equation and evolve $\Gamma_\Lambda$ from $\Lambda$ down to an
effective scale $k$ and observe whether or not the $\beta$-functions of the couplings between physical and gauge degrees
 of freedom are positive for a finite range of couplings. Since the exact renormalization group for the effective average
 action requires us to consider the most general action that is compatible with our field content and symmetries, we
 have to consider a truncation that contains all couplings that are ``low dimensional''. Similar ideas have been investigated in
\cite{Iliopoulos:1980zd}\cite{Alexandre:2009sy}\cite{Chadha:1982qq}.

Let us consider an ultralocal toy example consisting of a real scalar multiplet $\{\phi_i\}_{i=1}^n$ with an ultralocal
gauge group acting as
\begin{equation}
  \phi_i(x) \mapsto \lambda(x) + \phi_i(x)~.
\end{equation}
Let us change to the variables
\begin{equation}
  \begin{array}{rcl}
    \varphi_i&=&\phi_i-\phi_{i+1} \,\,\,\textrm{ for } 1 \le i < n~,\\
    \Phi &=& \sum_{i=1}^n\phi_i~,
  \end{array}
\end{equation}
where $\{\varphi_i\}_{i=1}^{n-1}$ are physical and $\Phi$ is pure gauge. To break this gauge invariance in a
controlled manner let us introduce an additional ``St\"uckelberg field'' $\chi$ transforming as
$\chi(x)\mapsto n\lambda(x) +\chi(x)$ under gauge transformations, so the most general local potential ansatz
with standard kinetic term allowed by gauge symmetry is
\begin{equation}
  \Gamma_k=\int d^4 x\left(\sum_{i=1}^{n-1}\frac 1 2 \partial_\mu \varphi_i \partial^\mu \varphi_i+\frac 1 2 \partial_\mu(\Phi-\chi)\partial^\mu(\Phi-\chi)+V_k(\varphi_i,\Phi-\chi)\right).
\end{equation}
Ultralocality allows us to fix a gauge $\chi\equiv 0$, making $\Phi$ physical in this gauge such that the original
gauge symmetry means that $\Phi$ decouples from the theory. Let us simplify the discussion by considering
$n=2$ and a truncation of the local potential to at most dimensionless couplings and let us impose an individual
$\mathbb Z_2$ symmetry, such that the truncation becomes:
\begin{equation}\label{eqn:toymodel}
  \Gamma_k=\int d^4x\left(\frac 1 2 \partial_\mu\varphi\partial^\mu\varphi+\frac 1 2 \partial_\mu\Phi\partial^\mu\Phi + \frac{k^2m^2_k}2 \varphi^2+\frac{k^2 M_k^2}2 \Phi^2+
  \frac{a_k}{4!}\varphi^4+\frac{b_k}{4}\varphi^2\Phi^2+\frac{c_k}{4!}\Phi^4\right).
\end{equation}
The $\beta$-functions can be extracted from (\ref{eqn:toytraces}):
\begin{subequations}
  \begin{flalign}
  \beta(M_k^2)&=-2M_k^2-\frac1{32\pi^2}\left(\frac{b_k}{(1+m_k^2)^2}+\frac{c_k}{(1+M_k^2)^2}\right)~,\\
  \nonumber \beta(b_k)&=\frac{1}{(4\pi)^2}\left(\frac{a_kb_k}{(1+m_k^2)^3}+\frac{b_kc_k}{(1+M_k^2)^3} \right.\\
   &\left.+\frac{2b_k^2}{(1+m_k^2)(1+M_k^2)^2}+\frac{2b_k^2}{(1+m_k^2)^2(1+M_k^2)}\right)~,\\
  \beta(c_k)&=\frac{3}{(4\pi)^2}\left(\frac{b_k^2}{(1+m_k^2)^3}+\frac{c_k^2}{(1+M_k^2)^3}\right)~.
  \end{flalign}
\end{subequations}
We see that the $\beta$-function of the dimensionless mass $M_k^2$ is always negative whereas the $\beta$-functions
of the four-point couplings $b_k$ and $c_k$ are always positive if $a_k,b_k,c_k>0$, meaning that the decoupling
limit $M_k$ large and $b_k\rightarrow 0$ is IR attractive in this regime. This implies that the renormalization group
flow drives towards original gauge symmetry, so no fine-tuning is necessary.

The opposite scenario is more familiar: If we consider a generic gauge theory, but forget about gauge symmetry,
then there are in general many low dimensional field monomials that are forbidden in the gauge theory that may
enter the theory without gauge symmetry. It is very unlikely that the unphysical degrees of freedom will decouple
from the physical degrees of freedom, since this requires that all new field monomials are RG irrelevant, whereas
low dimensional field monomials have the tendency to be RG relevant.

The most robust predictions of Wilsonian renormalization come from the discussion of fixed-points, which
 corresponds to the determination of universality classes. The IR-symmetry scenario can be seen at a glance when
 one considers the Gaussian fixed-point at which the masses are relevant and the four point couplings are
 (marginally) irrelevant. The IR-symmetries in the vicinity of any fixed-point are given by the symmetries of the
 relevant, marginally relevant and completely marginal directions at the fixed-point. Notice that i.g.~the IR-symmetry
may be affected by a nontrivial wave function renormalization. However, this is not the case if we are dealing with a Gaussian
fixed-point due to the following argument:
 If we introduce wave function renormalizations $Z_1,Z_2$
 in equation (\ref{eqn:toymodel}) then the $\beta$-function of the symmetry breaking coupling $B=Z_1 Z_2 b$ is
 $\beta(B)=(\eta_1+\eta_2)B+Z_1Z_2\beta(b)$, where $\eta_i=\partial_t Z_i$.
We obtain that $B$ has a Gaussian fixed-point if and only if $b$ has one and that $\partial_B\beta(B)=\partial_b\beta(b)$ at the
Gaussian fixed-point $b=B=0$.
Note that the stability of a non-Gaussian fixed-point may be affected by $\eta_i$. The running of $Z_i$ was however minimized
in test models by using the optimized cut-off \cite{Litim:2001dt}, so to gain a first insight, one may neglect the
effect of $\eta_i$, even for the discussion of generic fixed-points.

\section{\label{sec:cuscuton}Explicit renormalization of the cuscuton model}

In this section we discuss the renormalization of an adapted version of the cuscuton model
\cite{Afshordi:2009tt,Afshordi:2006ad} coupled to the Einstein-Hilbert truncation of general relativity in order
 to investigate the quantum behavior of the $K^2$ term in the Lorentz violating action (\ref{eqn:LVaction}).
  We make a general $\mathbb{Z}_2$-symmetric power-series potential ansatz for the effective action
\begin{flalign}
\label{eqn:EHcusc}
\Gamma_k[\bar g,h,\phi] = \Gamma_\mathrm{EH}[\bar g,h]  +  \int d^4 x \sqrt{g} \left( \sqrt{g^{\mu\nu}\phi_{,\mu} \phi_{,\nu}+\lambda^2}   + \sum\limits_{n=1}^{\infty}\frac{a_n}{(2n)!}  \phi^{2n}   \right)~,
\end{flalign}
including the quadratic cuscuton term (\ref{eqn:cuscaction}) for $n=1$.
  Note that, in order to have a background
independent FRGE, we introduced $\lambda=k^4\tilde \lambda$ with an arbitrarily small $\tilde\lambda$ in the kinetic term.
This parameter is treated as an external parameter and defines a family of quantum field theories. We show that the limit
 $\tilde\lambda\to0$ is well defined after the FRGE has been derived, leading to the cuscuton model.
The sole purpose of this parameter is to regularize the cuscuton propagator without introducing a cuscuton background,
as it will become clear below.
We will treat gravity within the Einstein-Hilbert truncation and focus on the renormalization
of the coupling constants $\kappa^2$, $\Lambda$ and $a_n$.

We start by collecting all terms required to evaluate the RHS of the flow equation (\ref{eqn:wetterich}).
The second variations $\Gamma^{(2)}_k$ are given in the appendix \ref{app:details}. Without loss of generality,
we may insert families of fields that project onto our truncation and which allow us to distinguish the field monomials therein.
For extracting the $\beta$-functions of the coupling constants $a_n$ it is therefore sufficient to insert a flat Euclidean
background metric $\bar g_{\mu\nu}(x)=\delta_{\mu\nu}$ as well as a constant cuscuton field $\phi(x)=\phi$.
Note that this is only a convenient choice that simplifies our calculations; the $\beta$-functions constructed therewith
 are generally valid.

As regulators $(R_k)_{AB}$ for the individual fields $A,B\in\lbrace \phi,h_{\mu\nu}\rbrace$ we use the properly
 normalized optimized regulator \cite{Litim:2001up} and define
\begin{subequations}
\label{eqn:regulators}
\begin{flalign}
(R_k)_{\phi\phi} &= \frac{1}{\lambda} (\bar \square+k^2) \Theta(1+\frac{\bar\square}{k^2})~,\\
 (R_k)_{hh}^{\mu\nu\rho\sigma} &= 2\kappa^2 K^{\mu\nu\rho\sigma} (\bar\square +k^2) \Theta(1+\frac{\bar\square}{k^2})~,
\end{flalign}
\end{subequations}
where all indices are raised and lowered with the background metric and where
\begin{flalign}
\label{eqn:gravitontensor}
 K^{\mu\nu\rho\sigma} = \frac{1}{4}\left( \bar g^{\mu\rho} \bar g^{\nu\sigma} +\bar g^{\mu\sigma} \bar g^{\nu\rho} -\bar g^{\mu\nu} \bar g^{\rho\sigma} \right)~.
\end{flalign}

The regularized second variations $(\Gamma^{(2)}_k +R_k)_{AB}$ can be split into a $\phi$ dependent part
$(F_k)_{AB}$ and $\phi$ independent part $(G_k)_{AB}$. The inverse is then given by the geometrical series
\begin{flalign}
 (\Gamma^{(2)}_k +R_k)^{-1}_{AB} =(P_k)_{AC}\cdot \sum\limits_{l=0}^{\infty} (-F_k\cdot P_k)^l_{CB}~,
\end{flalign}
where $(P_k)_{AB}:=(G_k)^{-1}_{AB}$ are the regularized propagators given by
\begin{subequations}
\label{eqn:propagators}
\begin{flalign}
(P_k)_{\phi\phi} &= \frac{\lambda }{k^2 +\lambda a_1}~,\\
 (P_k)_{hh}^{\mu\nu\rho\sigma} &= \frac{4 K^{\mu\nu\rho\sigma}}{2 \kappa^2 (k^2 - 2\Lambda - \frac{\lambda}{2\kappa^2})}~.
\end{flalign}
\end{subequations}
For $k^2\neq 0 $ and $k^2-2\Lambda\neq0$ we can always choose $\lambda$ sufficiently small in order to expand
 (\ref{eqn:propagators}) in $\lambda$. Thus the propagators are power-series in $\lambda$ with non-negative powers.

It turns out that the cuscuton does not contribute to the flow of the coupling constants $a_n$ in the limit $\lambda\to0$,
since all internal cuscuton propagators come with $\lambda^1$ and all field terms $(F_k)_{AB}$ come with non-negative
 powers of $\lambda$ as well. Even traces including $(\dot R_k)_{\phi\phi}\sim\lambda^{-1}$ do not contribute
 to the flow of the cuscuton potential, since they occur in the following combination
\begin{flalign}
 \mathrm{Tr}\left[(\dot R_k)_{\phi\phi} \cdot (P_k)_{\phi\phi}\cdot (F_k)_{\phi A}\cdot ~\dots~\cdot (P_k)_{\phi\phi}\right]=\mathcal{O}(\lambda^1)~.
\end{flalign}
The cuscuton only contributes to the flow of the gravitational sector, i.e.~$\kappa$ and $\Lambda$, via the trace
\begin{flalign}
\label{eqn:ehcontribution}
 \frac{1}{2}\mathrm{Tr}\bigl[(\dot R_k)_{\phi\phi} \cdot (P_k)_{\phi\phi}\bigr] =\mathcal{O}(\lambda^0)~.
\end{flalign}
In the limit $\lambda\to0$ this contribution to the renormalization of $\kappa$ and $\Lambda$ is equal to the
contribution of a canonical massless scalar field and does not depend on the cuscuton potential.
Thus we find that in the limit $\lambda\to0$ only gravity contributes to the $\beta$-functions of $a_n$. In particular,
only one term is contributing to the $\beta$-function of the quadratic term $a_1$
\begin{flalign}
 -\frac{1}{2}\mathrm{Tr}\left[(\dot R_k)_{hh} \cdot (P_k)_{hh}\cdot (F_k)_{hh}\cdot (P_k)_{hh}\right]~,
\end{flalign}
with the field term
\begin{flalign}
 (F_k)_{hh}= -a_1 K^{\mu\nu\rho\sigma} \frac{\phi^2}{2!} + \mathcal{O}(\phi^4)~.
\end{flalign}

This leads to the dimensionless $\beta$-function for $\tilde a_1=k^2 a_1$:
\begin{equation}
\beta(\tilde a_1) = \tilde a_1 \left( 2 + \frac{5}{192\pi^2\tilde\kappa^4} \frac{\beta(\tilde\kappa^2)+8\tilde \kappa^2}{ \left(1-2\tilde\Lambda\right)^2}\right)~,
\end{equation}
where we introduced the dimensionless coupling constants $\tilde \lambda_i$ defined by
$\lambda_i = k^{\mathrm{dim}\lambda_i} \tilde\lambda_i$.
The Lorentz violation is governed by the parameter $b=(4 a_1\kappa^2)^{-1}$, see (\ref{eqn:mum}).
Its $\beta$-function is
\begin{flalign}
\label{eqn:betab}
  \beta(b) = -b\left(\frac{\beta(\tilde\kappa^2) + 2\tilde\kappa^2}{\tilde\kappa^2}+\frac{5 }{ 192\pi^2 \tilde\kappa^4 } ~\frac{\beta(\tilde\kappa^2) +8\tilde\kappa^2}{\left(1-2\tilde\Lambda\right)^2}\right)~.
\end{flalign}
The $\beta$-functions of $\kappa^2$ and $\Lambda$ are equal to Reuter's $\beta$-functions
\cite{Reuter:1996cp} plus the additional contribution of a massless free scalar due to (\ref{eqn:ehcontribution}),
 which can e.g.~be found in \cite{Codello:2008vh}\cite{Benedetti:2009gn}.

We observe that the $\beta$-function of $b$ has a Gaussian fixed-point. Assuming real fixed-point values
$\kappa_*,\Lambda_*$ we see that $\partial_b \beta(b)<0$, meaning that the $(b=0,~\kappa=\kappa_*,~\Lambda=\Lambda_*)$
fixed-point is UV attractive.
Assuming a not necessarily Gaussian fixed-point $\kappa_*,\Lambda_*$, there appears to be the possibility for $\beta(b)$
to vanish for arbitrary values of $b$ due to vanishing of the bracket in (\ref{eqn:betab}). This bracket can however not
vanish for real values of $\kappa_*,\Lambda_*$, meaning that the only physically acceptable fixed-point for $b$ is
$b=0$. This means that the scenario of IR-attractivity of Lorentz symmetry, as explained in section \ref{sec:general},
is not realized in the cuscuton model. This does however not mean that this scenario may not be realized in models
with a different implementation of the Lorentz violation.

If we make the simplifying assumption that the $\beta$-functions of $\tilde\kappa^2$ and $\tilde\Lambda$ behave
classical in some finite interval of the scale $k$, meaning they have no anomalous scaling
\begin{flalign}
 \beta(\tilde\kappa^2)=-2\tilde\kappa^2~,\quad \beta(\tilde \Lambda) =-2\tilde\Lambda~,
\end{flalign}
then we obtain for (\ref{eqn:betab}) \begin{flalign}
\label{eqn:betabsimple}
 \beta(b)= - \frac{5 b e^{6t}}{32 \pi^2 \tilde\kappa_0^2 (e^{2t} -2\tilde\Lambda_0)^2}~,
\end{flalign}
where $t=\ln k$ and $\tilde\kappa_0,~\tilde\Lambda_0$ denote the ``initial conditions'' at $t=0$.
The physics described by this $\beta$-function (\ref{eqn:betabsimple}) is as follows: The absolute value of the
Lorentz violating parameter grows in the IR, i.e.~as $t\to-\infty$. This increase is bounded due to an exponentially
decreasing $\beta$-function. Thus one can in principle force $b$ to experimentally valid values by a suitable fine-tuning,
but the phenomenologically optimal scenario of a highly Lorentz violating UV theory which flows down to a sufficiently
Lorentz invariant IR theory can not be realized without seemingly unnatural assumptions or an unexpectedly large wave function renormalization, which we find unlikely due to our use of the optimized cut-off \cite{Litim:2001dt}.
The practical calculation of such a wave function renormalization would however require new techniques for the evaluation of the traces, since the cuscuton kinetic term can not be generated with the derivative expansion nor with a vertex (number of fields) expansion.

\section{\label{sec:matter}Adding matter fields}

Since the renormalization of the quadratic cuscuton term coupled to pure gravity obtained in the previous section does
not lead to IR attractivity of Lorentz symmetry, we will now study whether this can be achieved in presence of additional
matter fields.
The simplest standard matter model is an additional scalar field $\chi$ with canonical kinetic and mass term, together
 with a coupling to the cuscuton field via the lowest-order interaction
\begin{flalign}
 S_{\mathrm{int}} = g_\mathrm{mat} \vol \chi^2\phi^2~.
\end{flalign}
Note that even if this interaction would not be present at some scale, it would be induced radiatively
via the graviton interaction at a different scale. It is thus natural to consider such an operator.
We find that one additional trace contributes to the flow of $b$, namely
\begin{flalign}
 -\frac{1}{2}\mathrm{Tr} \left[(\dot R_k)_{\chi\chi}\cdot (P_k)_{\chi\chi}\cdot (F_k)_{\chi\chi}\cdot (P_k)_{\chi\chi}\right]~.
\end{flalign}
The terms involved in this trace are given by
\begin{subequations}
 \begin{flalign}
  (R_k)_{\chi\chi} &= (\bar\square +k^2)\Theta(1+\frac{\bar\square}{k^2})~,\\
  (P_k)_{\chi\chi} &= \frac{1}{k^2 +m_\chi^2}~,\\
  (F_k)_{\chi\chi} &= 2g_\mathrm{mat} \phi^2~.
 \end{flalign}
\end{subequations}
and the resulting $\beta$-function for $b$ is given by
\begin{flalign}
\label{eqn:betabmat}
  \beta(b) = -b\left(  \frac{\beta(\tilde\kappa^2) + 2\tilde\kappa^2}{\tilde\kappa^2} +\frac{5}{ 192\pi^2 \tilde\kappa^4 } ~\frac{\beta(\tilde\kappa^2) +8\tilde\kappa^2}{\left(1-2\tilde\Lambda\right)^2} -\frac{b \tilde g_\mathrm{mat} \tilde\kappa^2}{2\pi^2 (1+\tilde m_\chi^2)^2}\right)~.
\end{flalign}
We observe that this $\beta$-function can in principle be positive for a suitable choice of $b$ and
$\tilde g_\mathrm{mat}$. Since we are phenomenologically interested in small Lorentz violation $|b|\ll 1$
and small matter couplings $\vert \tilde g_\mathrm{mat}\vert \ll 1$ (the $K^2$ interpretation is invalid at
the scale at which the dimensionless matter couplings are not small), we obtain that matter i.g.~does not change
the sign of the $\beta$-function in our parameter range. On the other hand (\ref{eqn:betabmat}) leads to a condition
 for a non-Gaussian fixed-point $b=b_*$
\begin{flalign}
  b_*= \frac{\pi^2(1+\tilde m_{\chi*}^2)^2}{\tilde g_{\mathrm{mat}*} \tilde\kappa_*^2}\left( 4+
  \frac{5 }{ 12 \pi^2 \tilde\kappa^2_* }\frac{1}{(1-2\tilde\Lambda_*)^2}\right)~.
\end{flalign}
This means that if one had found a gravity-matter system with non-Gaussian fixed-point, then there could be
a non-Gaussian fixed-point for the gravity-matter-cuscuton system. However, even if one found such a system,
one would still generically have a negative $\beta$-function in the phenomenologically required regime, where
$|b|\ll 1$ as well as $|\tilde g_\mathrm{mat}|\ll 1$.

Similar effects can be obtained by introducing other matter fields, if they give rise to positive contributions to the
$\beta$-function of $b$. In particular, all matter effects will enter with $b^2$ and will be required to couple weakly
to the cuscuton field, so the $K^2$ interpretation holds.

\section{\label{sec:conc}Conclusions and outlook}

Motivated by the possibility of constructing UV complete theories of quantum gravity in explicitly Lorentz violating
theories with preferred foliation, we investigated the low energy effect of a preferred foliation in pure quantum
gravity. The particular model that we investigated was a version of cuscuton theory that has recently been shown
to be on-shell equivalent to  \hor -theory at low energies. We used a truncation ansatz consisting of Einstein-Hilbert
gravity coupled to a cuscuton field with local $\mathbb Z_2$-symmetric potential in Wetterich's exact renormalization
group equation and applied Reuter's background field methods for gravity. We were particularly interested in the
 nonperturbative and background independent renormalization of the quadratic cuscuton term which is related
 to the Lorentz violation. The nondynamical nature of the cuscuton field leads to very stable predictions for the
 renormalization of the cuscuton potential, which can be traced to the fact that only gravity propagates. We find in
 particular:
\begin{enumerate}
  \item The $\beta$-function of the quadratic cuscuton coupling is universal in the sense that it is independent
  of the rest of the cuscuton potential parameters.
  \item The only fixed-point of the Lorentz violating parameter is a UV-attractive Gaussian fixed-point.
  \item The inclusion of generic scalar matter has very weak effect on the UV-attractivity of the Gaussian fixed-point.
  \item Although having a quadratic term, the effect of the cuscuton on the renormalization of Newton's constant
  and the cosmological constant is equivalent to the effect of a massless scalar field.
\end{enumerate}
The physical interpretation of these results is that there might be some Lorentz violation at very low energy scales,
such as the Hubble scale, that may lead to observable cosmological effects. But using current cosmological bounds
 on Lorentz violation at the Hubble scale we expect no observable Lorentz violation in the high energy regime. This
 result contradicts the optimal scenario in which the renormalization group drives an initially large Lorentz violation
 in the UV theory to an unobservable value in the IR, hence if Lorentz violation is necessary to the UV theory one
 needs (within the universality class of the cuscuton model) an additional mechanism to tame this Lorentz violation in the IR or an unexpectedly large wave function renormalization.

To gain further insight into the effect of preferred foliations one has to investigate whether different gravity
models with preferred foliations lie in the same universality class as the cuscuton field. Within the cuscuton
theory itself one could gain further insight by performing calculations in truncations beyond the local potential ansatz, in particular the wave function renormalization. The investigation of the cuscuton wave function renormalization can not be calculated with standard techniques, because it is neither accessible in a derivative expansion nor in a vertex expansion.

\section*{Acknowledgements}
The authors thank Thorsten Ohl for valuable comments.
AS wants to thank the Perimeter Institute for Theoretical Physics for the warm hospitality.
This research was supported by Perimeter Institute for Theoretical Physics. Research at Perimeter Institute for
Theoretical Physics is supported in part by the Government of Canada through NSERC and by the Province of
Ontario through MRI. AS is supported by Graduiertenkolleg 1147 ``Theoretical Astrophysics and Particle Physics''.
\begin{appendix}

\section{\label{app:beta}Toy model calculation}

Using the same regulator $R_k$ for all fields, the fluctuation matrix $\Gamma^{(2)}_k+R_k$ of
(\ref{eqn:toymodel}) can be split into $F+G$ with:
\begin{subequations}
\begin{flalign}
  G&=\left(\begin{array}{cc}
    R_k+p^2+k^2m_k^2 & 0\\
    0 & R_k+p^2+k^2M_k^2
  \end{array}\right)~,\\
  F&=\left(\begin{array}{cc}
    \frac{a_k}2 \varphi^2 +\frac{b_k}{2} \Phi^2& b_k \varphi\Phi\\
    b_k \varphi\Phi & \frac{b_k}{2}\varphi^2+\frac{c_k}2 \Phi^2
  \end{array}\right).
\end{flalign}
\end{subequations}
To project the RHS of Wetterich's equation onto the local potential ansatz, we may insert constant fields,
and extract the $\beta$-functions for the masses from the trace $-\frac 1 2 \textrm{Tr}(P\dot RPF)$
and for the four-point couplings from $\frac 1 2 \textrm{Tr}(P\dot RPFPF)$ respectively, where $P=G^{-1}$.
Using the optimized cut-off $R_k=(k^2-p^2)\Theta(1-\frac{p^2}{k^2})$ we find for constant fields:
\begin{flalign}\label{eqn:toytraces}
  \begin{array}{rcl}
    -\frac 1 2 \textrm{Tr}(P\dot RPF)&=&-\frac{k^2}{64\pi^2}\int d^4x\left(\left(\frac{a_k}{(1+m_k^2)^2}+\frac{b_k}{(1+M_k^2)^2}\right) \varphi^2+\left(\frac{b_k}{(1+m_k^2)^2}+\frac{c_k}{(1+M_k^2)^2}\right) \Phi^2\right)\\
    \frac 1 2 \textrm{Tr}(P\dot RPFPF)&=& \frac{1}{64 \pi^2}\int d^4x\left(\frac{(a_k\varphi^2+b_k\Phi^2)^2}{2(1+m_k^2)^3}+\frac{2 b_k^2 \varphi^2\Phi^2}{(1+m_k^2)^2(1+M_k^2)} +\frac{2b_k^2\varphi^2\Phi^2}{(1+m_k^2)(1+M_k^2)^2}+\frac{(b_k \varphi^2+c_k \Phi^2)^2}{2(1+M_k^2)^3}\right)
  \end{array}
\end{flalign}

\section{\label{app:details}Second variation of the cuscuton model}
We give the complete second variation of the effective action (\ref{eqn:EHcusc}). It is obtained by plugging
in the background field expansion $g_{\mu\nu}=\bar g_{\mu\nu}+\epsilon h_{\mu\nu}$ and
$\phi=\bar\phi+\epsilon f$ into the effective action (\ref{eqn:EHcusc}) and collecting the $\epsilon^2$
 terms\footnote{Notice that this procedure also produces the second variation w.r.t.~fields not having backgrounds,
 because the second variation can be obtained as the second variation with respect to the fluctuations of a
 functional of background field plus fluctuation evaluated at vanishing background.}.
For a better readability we give independently the individual parts of the second variation.

The second variation of the kinetic cuscuton action is given by
\begin{multline}
\Gamma^{(2)}_\mathrm{kin} = \volbar \left\lbrace  \left(\frac{h^2}{8} -\frac{h^{\mu\nu}h_{\mu\nu}}{4}\right) \sqrtbar   - \frac{h}{4} ~\frac{h^{\mu\nu}\partial_\mu \bar\phi\partial_\nu\bar\phi - 2\partial^\mu\bar\phi\partial_\mu f}{\sqrtbar}\right.   \\
\left.+\frac{1}{2} \frac{h^{\mu\lambda}h_{\lambda}^{~\nu}\partial_\mu\bar\phi \partial_\nu\bar\phi - 2h^{\mu\nu}\partial_\mu\bar\phi \partial_\nu f + \partial^\mu f\partial_\mu f}{\sqrtbar} - \frac{1}{8} \frac{(h^{\mu\nu}\partial_\mu\bar\phi\partial_\nu\bar\phi -2\partial^\mu\bar\phi\partial_\mu f  )^2 }{\sqrtbar^{~3}}\right\rbrace~.
\end{multline}
The second variation of the cuscuton potential part is
\begin{flalign}
\Gamma^{(2)}_\mathrm{pot} = \sum\limits_{n=1}^{\infty}\volbar~a_n \left\lbrace \left(\frac{h^2}{8}  -\frac{h^{\mu\nu}h_{\mu\nu}}{4} \right)    \frac{\bar\phi^{2n}}{(2n)!} +\frac{h f}{2}  \frac{\bar\phi^{2n-1}}{(2n-1)!}   +\frac{f^2}{2}  \frac{\bar\phi^{2n-2}}{(2n-2)!}   \right\rbrace ~.
\end{flalign}
\end{appendix}
The second variation of the Einstein-Hilbert part can be found in \cite{Reuter:1996cp} for  general
background metric fields. Insertion of a flat background metric into the general variation yields:
\begin{flalign}
 \Gamma^{(2)}_\mathrm{grav} = -\kappa^2 \int d^4x~ h_{\mu\nu} K^{\mu\nu\rho\sigma} (\square +2\Lambda) h_{\rho\sigma}~,
\end{flalign}
where $K^{\mu\nu\rho\sigma}$ was defined in (\ref{eqn:gravitontensor}).

The functional derivatives with respect to the individual fluctuation fields $\lbrace f,h_{\mu\nu}\rbrace$ of
the sum of the expressions above enter the RHS of the FRGE (\ref{eqn:wetterich}).

\end{document}